# Bigravity as an interpretation of the cosmic acceleration

**Jean-Pierre Petit ( phd, astrophysicist, retired )**

**and Gilles d'Agostini, phd**

jppetit1937@yahoo.fr

___

**Abstract**:

A bimetric description of the Universe is proposed, which may explain the cosmic acceleration deduced from optically observed high-redshift supernovae. This phenomenon is classically charged to a mysterious repulsive "dark energy" whose true nature remains unknown. It is identified in the present model with a set of particles having negative mass and energy. These negative mass particles are emitting negative energy photons, hence they are not observable using optical instruments.

___

## 1 – Introduction

For years, astrophysicists and cosmologists have tried to define which of the three Friedmann models, solutions to Einstein's equations, fit the observations. With time, a general consensus had been established to conclude that in Einstein's equations, the cosmological constant $\Lambda$ had to be zero. Thus the cosmological model was supposed to rise from the Friedmann equation:
(1)
$$R^2 R'' + a^2 = 0$$

where the second derivative of $R(t)$ could only be negative. However, measurements performed on distant supernovae over the last decade have revealed an unexpected acceleration of the cosmic expansion, $R''$ becoming positive for high redshifts[, deep space]. Todays, specialists think that cosmic dynamics depends on three components:

- Visible matter: 4 %
- Dark matter (attractive): 22 %
- Dark energy (repulsive): 74 %

The last component, whose existence had been neglected until now recently, might well explain the observed acceleration if it is basically related to *repulsive* forces.

As shown below, it is possible to obtain such an effect if we suppose the Universe consisting of:

- Particles with non-zero, positive mass and positive energy.

- Particles with zero mass: positive energy photons.

- Particles with non-zero, negative mass and negative energy.

- Particles with zero mass: negative energy photons.

Group theory accounts for this since a long time. Relativistic dynamics of a material point is associated with the dynamic Poincaré group, which is based on the Lorentz group. If $\Gamma$ is the Gramm matrix:

(2)
$$\Gamma = \begin{bmatrix} 1 & 0 & 0 & 0 \\ 0 & -1 & 0 & 0 \\ 0 & 0 & -1 & 0 \\ 0 & 0 & 0 & -1 \end{bmatrix}$$

the Lorentz group can be represented by (4,4) matrices $L$. Hence:

(3)
$$^t L \, \Gamma \, L = \Gamma$$

If $C$ is the space-time translation vector:

(4)
$$C = \begin{bmatrix} \Delta t \\ \Delta x \\ \Delta y \\ \Delta z \end{bmatrix}$$

then Poincaré's group corresponds to the matrix:

(5)
$$\begin{bmatrix} L & C \\ 0 & 1 \end{bmatrix}$$

The Lorentz group owns four connected components. Using a terminology borrowed from reference [1] the first two ones form the *orthochron* subgroup and do not reverse the time coordinate. The last two ones constitute the *antichron* subset and their components reverse time. The Poincaré group, owing ten dimensions, inherits this property and also owns four connected components. Let us call $L_o$ the matrices of the *antichron subgroup* of the Lorentz group. The set of matrices made from the neutral component $L_n$ of the Lorentz group is traditionally known as the *restricted* Poincaré group:
(5bis)
$$\begin{bmatrix} L_n & C \\ 0 & 1 \end{bmatrix}$$

Now, writing the matrix of the orthochron subgroup of Poincaré's group:
(6)
$$\begin{bmatrix} L_o & C \\ 0 & 1 \end{bmatrix}$$

the *complete* group can be written:
(7)
$$\begin{bmatrix} \lambda L_o & C \\ 0 & 1 \end{bmatrix} \quad \text{with } \lambda = \pm 1$$

In reference [1] J.M.Souriau built the coadjoint action of Poincaré's group on its momentum space, which owns the same number of dimensions as the group. He writes the momentum by forming, with four of its components, the energy-impulsion vector $P$ :
(8)
$$P = \begin{bmatrix} E \\ p_x \\ p_x \\ p_x \end{bmatrix}$$

and bringing together the last six components, which form an antisymmetric matrix $M$, Souriau then writes the momentum of Poincaré's group:
(9)
$$J = \{ M , P \}$$

The coadjoint action of Poincaré's group on its Lie algebra then gives:
(10)
$$P' = \lambda L_o P$$

(11)
$$M' = L_o M\, {}^tL_o + ( C\, {}^tP\, {}^tL_o - L_o P\, {}^tC )$$

This result can be found in reference [1], page 198. It clearly shows that the antichron components of Poincaré's group transform any *orthochron* movement (*corresponding to a {M,P} set of the momentum components*) with positive energy into an *antichron* movement with *negative energy*. As highlighted by the author of "Structure of Dynamical Systems", in order to avoid the problems caused by negative energy (and negative mass) particles, relativistic dynamics of material points can arbitrarily be limited to the restricted Poincaré group.

## 2) A world populated by positive and negative energy particles

But todays, particles with negative energy and mass should be taken into account, which implies having recourse to the *complete* Poincaré group, outfitted with its orthochron and antichron components.

Notice importantly that antichron components of the group also transform orthochron photons movements with positive energy into antichron movements with negative energy.

We can then suppose that real observers, or measuring instruments, being made of particles with positive mass and energy, they are not able to deal with negative energy photons, and conversely that hypothetic observers and measuring instruments made of negative mass and energy particles, would not be able to deal with positive energy photons.

From this point of view the two cosmic contents, one with positive energy and the other with negative energy, cannot interact through electromagnetic waves. Concretely, this implies that if our Universe contains negative mass and energy particles, we cannot see them nor record images of them with optical devices such as telescopes, whatever the frequency considered.

## 3) The true nature of antimatter

It is possible to consider that antimatter particles correspond to other kinds of a material point's relativistic movements. The electric charge $q$ can be geometrized, making of it an additional component of the momentum, by extending Poincaré's group. Introducing an additional parameter $\nu$ we can then bring out (as explained in reference [2], chapter 5, page 413) the fact that the fifth dimension of Kaluza comes with a charge reversal. Let us introduce the four-vector:

(12)
$$\xi = \begin{bmatrix} t \\ x \\ y \\ z \end{bmatrix}$$

and the five-vector:
(13)
$$\begin{bmatrix} \varsigma \\ \xi \end{bmatrix}$$

where $\varsigma$ represents the additional dimension of Kaluza. We define this property by introducing the group of four connected components:

(14)
$$\begin{bmatrix} \varsigma' \\ \xi' \\ 1 \end{bmatrix} = \begin{bmatrix} \nu & 0 & \nu\phi \\ 0 & L_o & C \\ 0 & 0 & 1 \end{bmatrix} \times \begin{bmatrix} \varsigma \\ \xi \\ 1 \end{bmatrix} = \begin{bmatrix} \nu\varsigma + \nu\phi \\ L_o\xi + C \\ 1 \end{bmatrix}$$

The group momentum has 11 components:

(15)
$$J = \{\, q\,,\, M\,,\, P\,\}$$

where $q$ is the electric charge. And the action of the group on its momentum space thus gives:
(16)
$$q' = \nu\, q$$

(17)
$$P' = L_o\, P$$

(18)
$$M' = \left(L_o\, M\,{}^t L_o\right) + (C\,{}^t P\,{}^t L_o) - (L_o\, P\,{}^t C)$$

Starting from the movement of a material point supposed equivalent to a matter particle of positive mass and energy, the action of an element ( $\nu = -1$ ) from the group will reverse the extra dimension of Kaluza and its associated charge, and this movement will become that of an antimatter particle. Since this action does not concern the four-vector $P$, it is classically deduced that antimatter has positive energy and mass.

3) **About negative energy antimatter**

If we decide that our physics, that relativistic dynamics of the material point, are dependent on the restricted dynamic Poincaré group or on the orthochron subgroup of the Poincaré group, then the Universe is only made of particles having both mass and energy of the same sign (positive).

If we extend the group as shown above, by introducing the additional dimension $\zeta$ and by doubling the number of components (which goes from two to four) we obtain a Universe populated by matter and antimatter.

If, in the group above, the Lorentz group restricted to its orthochron components is replaced by the complete Lorentz group, we obtain a group of eight connected components with six kinds of particles:

- *Matter particles of positive energy and mass.*
- *Antimatter particles of positive energy and mass.*
- *Photons of positive energy (the photon is its own antiparticle)*

- *Matter particles of negative energy and mass.*
- *Antimatter particles of negative energy and mass.*
- *Photons of negative energy (the photon is its own antiparticle)*

By combining in the following way (admittedly arbitrary) the transformations reversing space, time and charge, we can write:
(19)
$$\begin{bmatrix} \lambda\nu & 0 & \lambda\nu\phi \\ 0 & \lambda L & C \\ 0 & 0 & 1 \end{bmatrix}$$

Then the action of the group on its momentum space is written:

$$q' = \lambda \, v \, q$$

$$P' = \lambda \, L_o \, P$$

$$M' = L_o \, M \, {}^t L_o \; + \; ( \, C \, {}^t P \, {}^t L_o \, - \, L_o \, P \, {}^t C \, )$$

Hence, reversing the sign of $\lambda$ brings a reversal of charge, space and time, in other words a *CPT-symmetry*.

4) **Construction of an interaction between positive energy particles and negative energy particles. Bimetric configuration, bigravity.**

By choosing the complete Poincaré group, we obtained an invisible « second matter species ». In order for it to have the properties of dark energy, laws of gravitational interaction giving a repulsive nature to this « matter-dark energy » are needed.

Writing Einstein's equation:
(20)
$$S = \chi \, T - g \, \Lambda$$

$S$ is a geometrical tensor, i.e. Einstein's tensor $S(g)$ built from the metric $g$ associated to a four-dimensional manifold. $T$ is a tensor density representing the energy-matter content of the Universe, $\chi$ is Einstein's constant:
(21)
$$\chi = - \frac{8 \, \pi \, G}{c^2}$$

and $\Lambda$ is the cosmological constant.

$g$ belongs to a function space sometimes called *hyperspace*. The Einstein tensor is, in absolute value, the gradient of the Hilbert functional:
(22)
$$H(g) = \int_M (a \, R(g) + b) \, dV_g$$

in which *a and b are constants and R the scalar curvature.* This functional is invariant by diffeomorphism. Einstein's constant $\chi = 1/a$ derives from constant *a* only, while *a* and *b* give the cosmological constant $\Lambda = b/2a$ . Let us set $\Lambda = 0$, which amounts to take $b = 0$. Thereafter we will not need it. So we restrict (22) to:
(23)
$$H(g) = \int_M a \, R(g) \, dV_g$$

We build the Einstein tensor $S$ as a gradient of this fonctional in the space of metrics. Associated with a tensor density $T$ we obtain the equation:
(24)

$$S = \chi T$$

where the unknown term is the metric *g* and where the problem directly hinge upon the value of this tensor density *T*.

Now we suppose the Universe is a four-dimensional manifold $M_4$ fitted up with *two metrics*, which we call $g^+$ and $g^-$.

We also suppose that positive energy particles follow the geodesics resulting from metrics $g^+$, while negative energy particles progress along geodesics resulting from metrics $g^-$.

It is geometrically impossible for a positive energy particle to move along a geodesic deduced from the metric $g^-$, and conversely for a negative energy particle to move along a geodesic deduced from the metric $g^+$.

From the metric $g^+$ a geometrical tensor $S^+$ can be built, and in the same manner from the metric $g^-$ a geometrical tensor $S^-$. Let us take the Einstein field equation again, which we will write:
(25)
$$S^+(g^+) = \chi T$$

*T* is a tensor density which contains the problem data, the energy-matter density at the considered point. This is the conjugation of two contributions.

$\rho^+$ will designate the density of positive energy matter.

In order to tally with previous works (references [7] to [11]) we decide to designate with $\rho^-$ the *absolute value* of the density of negative mass matter. Same thing for the pressure terms involved in the tensor density of this second kind of matter, so we will designate with $T^-$ the *absolute value* of the tensor density of an energy-matter corresponding to negative energy and mass. Thereby we write:
(26)
$$S^+(g^+) = \chi(T^+ - T^-)$$

Now we are going to create a coupling between both metrics, and we will call them *conjugate metrics*. Let us take again the development which brought us to the construction of the Einstein's equation. We write the Hilbert functional:
(27)
$$H^+(g) = \int_M a R^+(g) \, dV_g$$

from where the derivative gives the geometrical tensor $S^+(g^+)$ which is the gradient of this functional in the metrics function space, in the hyperspace. We suppose the geometrical tensor

$$S^-(g^-)$$

is a derivative from the Hilbert functional:
(28)

$$H^-(g) = \int_M a R^-(g) \, dV_g = -H^-(g)$$

Since the geometrical tensor is the gradient of the functional in the metrics space, this implies
(29)
$$S^+(g^+) = -S^-(g^-)$$

and hence both metrics are not independant.

Let us call them *conjugate metrics,* a new mathematical concept.

If metric $g^+$ is known, then metric $g^-$ can be deduced. We can write the two equations (26) and (29) as follow:
(30)
$$S^+(g^+) = \chi(T^+ - T^-)$$
(31)
$$S^-(g^-) = \chi(T^- - T^+)$$

And it is in that form this coupled field equations system had been published for the first time in the *Nuevo Cimento* review in 1994 ([10]).

5) **Newtonian approximation**

This approximation allows to understand the dynamics arising from this system. We will expand these equations into a series, on the basis of a situation at zero order, where the second members are null and where the metrics, assumed as stationary, are Lorentzian. We suppose curvatures are weak and agitations velocities in the two « cosmic fluids » are low compared to the speed of light. We have:
(32)
$$g^+ = \eta^+ (\text{Lorentz}) + \varepsilon \, \gamma^+$$
(33)
$$g^- = \eta^- (\text{Lorentz}) + \varepsilon \, \gamma^-$$

where $\varepsilon$ is a small parameter. We expend the two metrics into series, using the traditional method. This calculation is present in all books dealing with General Relativity (i.e. in [the] reference [12], chapter 4, section 4.3). As a result, we get the following for the movements of positive mass and energy particles, along geodesics of the metric $g^+$:
(34)
$$\frac{d^2 x^i}{dt^2} = -c^2 \varepsilon \frac{\partial \gamma^{+oo}}{\partial x^i}$$

and for the movements of negative mass and energy particles, along geodesics of the metric $g^-$:
(35)
$$\frac{d^2 x^i}{dt^2} = -c^2 \varepsilon \frac{\partial \gamma^{-oo}}{\partial x^i}$$

This leads to define the gravitational potentials regulating these movements:

(36)
$$\Psi^+ = \frac{c^2}{2} \varepsilon \gamma^{+oo}$$

(37)
$$\Psi^- = \frac{c^2}{2} \varepsilon \gamma^{-oo}$$

In the Newtonian approximation, the tensor density is reduced to a single term $T^{oo}$

(38)
$$T = T^+ - T^- \cong \begin{bmatrix} (\rho^+ - \rho^-) & 0 & 0 & 0 \\ 0 & 0 & 0 & 0 \\ 0 & 0 & 0 & 0 \\ 0 & 0 & 0 & 0 \end{bmatrix}$$

Series developments give:

(39)
$$-\varepsilon \left( \frac{\partial^2 \gamma^{+oo}}{\partial x^2} + \frac{\partial^2 \gamma^{+oo}}{\partial y^2} + \frac{\partial^2 \gamma^{+oo}}{\partial z^2} \right) = \chi ( \rho^+ - \rho^- )$$

(40)
$$-\varepsilon \left( \frac{\partial^2 \gamma^{-oo}}{\partial x^2} + \frac{\partial^2 \gamma^{-oo}}{\partial y^2} + \frac{\partial^2 \gamma^{-oo}}{\partial z^2} \right) = \chi ( \rho^- - \rho^+ )$$

Taking into account (36), (37) we find two coupled Poisson's equations:

(41)
$$\frac{\partial^2 \Psi^+}{\partial x^2} + \frac{\partial^2 \Psi^+}{\partial y^2} + \frac{\partial^2 \Psi^+}{\partial z^2} = 4 \pi G ( \rho^+ - \rho^- )$$

(42)
$$\frac{\partial^2 \Psi^-}{\partial x^2} + \frac{\partial^2 \Psi^-}{\partial y^2} + \frac{\partial^2 \Psi^-}{\partial z^2} = 4 \pi G ( \rho^- - \rho^+ )$$

Which gives us the extremely simple relation:

(43)
$$\Psi^- = - \Psi^+$$

Following : how positive and negative mass particles interact:

- Positive mass particles attract each others through Newton's law
- Negative mass particles attract each others through Newton's law
- Particles having mass of opposite sign repel each others through « anti-Newton law »

This result could also have been obtained starting from the expression:
(44)
$$F = -\frac{G\,m\,m'}{d^2}$$

One advantage of such a description of the « dark energy » is that its nature becomes rather simple: its components are merely the same as those of normal matter, but simply with negative mass and energy. They are no longer « exotic » but consist of:

- *Protons and antiprotons with negative mass and energy*
- *Neutrons and antineutrons with negative mass and energy*
- *Electrons and anti-electrons with negative mass and energy*
- *Photons with negative energy*

6) **About the cosmic acceleration**

In cosmology, observations show a reacceleration of the Universe for high redshifts. Can we reach such a result by identifying the dark energy to this negative energy-matter? To do so a cosmological model will be built, based on a solution in the form of a couple of conjugated metrics ( $g^+$, $g^-$ ). We take riemannian metrics of signature ( + - - - ) adding the hypothesis of homogeneity and isotropy. This leads us to Robertson-Walker metrics:
(45)
$$ds^2 = c^2\,dt^2 - R_{(t)}^{+\;2}\,\frac{1}{1+\frac{k}{4}u^2}\,(du^2 + u^2\,d\theta^2 + u^2\,\sin^2\theta\,d\varphi^2)$$

Our Universe is *bimetric*. Points of this manifold can be located using an arbitrary frame of reference. In the above expression we find

- *a time-marker t*
- *space markers u, θ, φ*

The first term designates the time coordinate, and the three others the space coordinates. Such quantities can be considered as simple numbers, or angles, and be written:

$$[\,\tau,\,u,\,\theta,\,\varphi\,]$$

Let us show that theses coordinates are a common frame of reference for two sets of mass-points corresponding to positive and negative energies. Consider two points A and B, with coordinates:

$$[\,u_A,\,\theta_A,\,\varphi_A\,] \quad \text{and} \quad [\,u_B,\,\theta_B,\,\varphi_B\,]$$

If these two sets of numbers remain constant through time, such points are *co-moving*. Since we are in a situation of isotropy and homogeneity, points A and B can be linked through geodesic curves, however calculated from the basically different metrics $g^+$ et $g^-$. Lengths may differ, insofar as their corresponding scale factors $R^+$ and $R^-$ might no more be equal.

In an expanding or simply evolving Universe, both scale factors will vary according to laws:

$$R^+(\tau) \quad \text{and} \quad R^-(\tau)$$

So we get two measurements of arc AB and two different lengths, whether the measurement is taken in the world of positive energies and masses with the metric $g^+$, or in the world of negative energies and masses with the metric $g^-$.

- According to the first measurement, the length of arc AB will vary like $R^+(\tau)$
- According to the second measurement, the length of arc AB will vary like $R^-(\tau)$

Now consider a 3-ball $B_3$, whose boundary is a sphere $S_2$. Once again we get different measurements for the surface of the sphere $S_2$ or the volume of the 3-ball $B_3$ depending on the metric we rely on :

- Refering to metric $g^+$ the area of sphere $S_2$ will vary like $R^{+2}$ and the volume of ball $B_3$ like $R^{+3}$

- Refering to metric $g^-$ the area of sphere $S_2$ will vary like $R^{-2}$ and the volume of ball $B_3$ like $R^{-3}$

$\rho^+$ and $\rho^-$ represent the densities of the two species. By expressing mass conservation, we write:
(46)
$$\rho^+ R^{+3} = cst$$

(47)
$$\rho^- R^{-3} = cst$$

Let us introduce two Robertson-Walker metrics into the field equations and suppose the Universe is only filled with dust, we will end to the following four equations system in which the pressure terms can be neglected:
(48)
$$\frac{3(R^{+'})^2}{c^2 R^+} + \frac{3k}{R^{+2}} = \frac{8\pi G}{c^2}(\rho^+ - \rho^-)$$

(49)
$$\frac{2 R^{+''}}{c^2 R^+} + \frac{(R^{+'})^2}{c^2 R^{+2}} + \frac{k}{R^{+2}} = 0$$

(50)
$$\frac{3(R^{-'})^2}{c^2 R^-} + \frac{3k}{R^{-2}} = \frac{8\pi G}{c^2}(\rho^- - \rho^+)$$

(51)
$$\frac{2 R^{-''}}{c^2 R^-} + \frac{(R^{-'})^2}{c^2 R^{-2}} + \frac{k}{R^{-2}} = 0$$

Taking into account mass conservation, we get :
(52)
$$R^{+2} R^{+\prime\prime} + 1 - \frac{R^{+3}}{R^{-3}} = 0$$

(53)
$$R^{-2} R^{-\prime\prime} + 1 - \frac{R^{-3}}{R^{+3}} = 0$$

a system with the trivial solution:
(54)
$$R^{+} = R^{-} = a t$$

which is unstable :

$$\text{If} \quad R^{+} > R^{-} \quad \text{then} \quad R^{+\prime\prime} > 0 \quad \text{and} \quad R^{-\prime\prime} < 0$$

and vice-versa.

We thus obtain an acceleration effect on the expansion for the positive mass and energy population, and a reverse phenomenon for the negative mass and energy population:

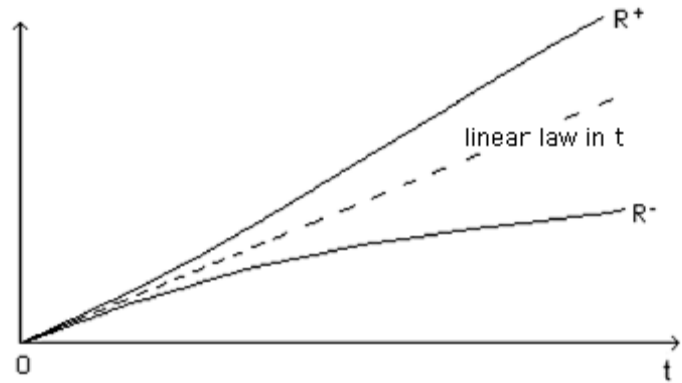

**Scale factors evolution**

We must remember that what we actually measure is a redshift related to the derivative of the scale factor $R$. The important thing is to take account of the observations. Measurements of the recession speed for the most remote supernovae have revealed a re-acceleration of the expansion that is commonly charged to a mysterious dark energy. The model presented here is an alternative interpretation for its nature.

**Conclusion**

The present work, based on a bimetric description of the Universe going with a bigravity dynamics, accounts for the re-acceleration of the part of the Universe accessible to optical observations, a phenomenon charged until now to a *mysterious repulsive dark energy* whose true nature remains unknown. Our model fully accounts for this phenomenon, through a new concept of negative mass and energy particles: emitting negative energy photons, they are not observable using optical instruments.

Applied to the very early Universe, our model provides a common law of linear evolution for the scale factors $R^+$ et $R^-$ with respect to time. Further improvements will be given in a forthcoming paper, including "variable constants" (previously introduced by the main author in 1988 in papers [13], [14], [15], then in 1995 in reference [9]).

**Appendix:** The program of the CITV 2007 Meeting.

1  Pascal Adjamagbo « *Symplectic geometry with positive characteristic* »

2  Kossi Atchonouglo « *Special, symmetric, tridiagonal matrix. Calculation of the eigenvalues and eigenvectors* »

3  Dominique Chevallier « *Holonomy group for dynamical systems* »

4  Dan Dimitriu « *Rigid bodies dynamics* »

5  Boris Kolev « *Introduction to Ricci flow* »

6  Boris Kolev « *Structure equations for Killing tensors* »

7  Jean-Pierre Petit « *Negative mass and energy. Complete Poincaré group* »

8  Jean-Pierre Petit « *MHD* : *The Z-machine* »

9  Jean-Pierre Petit « *The trouble with astrophysics* »

10  Michel Petitot « *Equivalence problems and groupoids* »

11  Géry de Saxcé « *Path integral* »

12 Gijs Tuynman  « *To cut and to glue* »